\documentclass[final,5p,times,twocolumn]{elsarticle}

\usepackage{adjustbox}
\usepackage{graphicx}
\usepackage{flushend} 
\usepackage{ccicons}
\usepackage{makerobust} 
\MakeRobustCommand\rotatebox 
\graphicspath{{./pdf/}{./jpeg/}{./images/}}
\DeclareGraphicsExtensions{.pdf,.jpeg,.jpg,.png}
\DeclareGraphicsExtensions{.pdf,.jpeg,.jpg,.png} 
\usepackage{amssymb}
\usepackage{pifont}

\usepackage{pseudocode}
\usepackage{verbatim}
\usepackage{sverb}

\usepackage{amsmath}
\usepackage[ruled]{algorithm2e}
\usepackage[noend]{algpseudocode}

\usepackage[nodots,nocompress]{numcompress}

\usepackage{textcomp}

\PassOptionsToPackage{hyphens}{url}
\usepackage[hidelinks]{hyperref}
\hypersetup{
    colorlinks,
    linkcolor=blue,
    citecolor=blue,
    urlcolor=blue
}

\usepackage{upgreek}

\usepackage{url}
\usepackage{float}

\usepackage{amsmath}

\usepackage[T1]{fontenc}

\usepackage{array}
\usepackage{booktabs}
\usepackage{xcolor}

\usepackage{units}

\usepackage{soul}
\usepackage{color}
	\definecolor{celadon}{rgb}{0.67, 0.88, 0.69}
  \definecolor{flamingopink}{rgb}{0.99, 0.56, 0.67}

\usepackage{makecell}

\usepackage{enumitem}

\usepackage{listings}

\usepackage{caption} 
\usepackage[labelformat=simple]{subcaption}

\biboptions{numbers}

\begin{document}

\begin{frontmatter}

\title{ChatGPT for Digital Forensic Investigation: The Good, The Bad, and The Unknown}

\author[add1]{Mark Scanlon\corref{firstcorr}}
\ead{mark.scanlon@ucd.ie}
\cortext[firstcorr]{Corresponding author}
\address[add1]{Forensics and Security Research Group, School of Computer Science, University College Dublin, Ireland}

\author[add2]{Frank Breitinger}
\ead{frank.breitinger@unil.ch}
\address[add2]{School of Criminal Justice, University of Lausanne, Lausanne, Switzerland}

\author[add3]{Christopher Hargreaves}
\ead{christopher.hargreaves@cs.ox.ac.uk}
\address[add3]{Department of Computer Science, University of Oxford, United Kingdom}

\author[add4]{Jan-Niclas Hilgert}
\ead{hilgert@cs.uni-bonn.de}
\address[add4]{Fraunhofer FKIE, Bonn, Germany}

\author[add5]{John Sheppard}
\ead{frank.breitinger@unil.ch}
\address[add5]{Department of Computing and Mathematics, South East Technological University, Waterford, Ireland}

\begin{abstract}
The disruptive application of ChatGPT (GPT-3.5, GPT-4) to a variety of domains has become a topic of much discussion in the scientific community and society at large. Large Language Models (LLMs), e.g., BERT, Bard, Generative Pre-trained Transformers (GPTs), LLaMA, etc., have the ability to take instructions, or prompts, from users and generate answers and solutions based on very large volumes of text-based training data. This paper assesses the impact and potential impact of ChatGPT on the field of digital forensics, specifically looking at its latest pre-trained LLM, GPT-4. A series of experiments are conducted to assess its capability across several digital forensic use cases including artefact understanding, evidence searching, code generation, anomaly detection, incident response, and education. Across these topics, its strengths and risks are outlined and a number of general conclusions are drawn. Overall this paper concludes that while there are some potential low-risk applications of ChatGPT within digital forensics, many are either unsuitable at present, since the evidence would need to be uploaded to the service, or they require sufficient knowledge of the topic being asked of the tool to identify incorrect assumptions, inaccuracies, and mistakes. However, to an appropriately knowledgeable user, it could act as a useful supporting tool in some circumstances.
\end{abstract}

\end{frontmatter}

\section{Introduction}
\label{intro}

The emergence of Generative Artificial Intelligence (GAI) has sparked significant interest and scrutiny across various disciplines, including its potential impact on scientific research and writing~\cite{DWIVEDI2023102642, alkaissi2023artificial,thorp2023ChatGPTNotAnAuthor}. In particular, Large Language Models (LLMs), such as ChatGPT -- released in November 2022 (\url{openai.com/blog/ChatGPT}), have been identified as having numerous beneficial use cases and risks in various fields including digital forensic (DF) investigation~\cite{scanlon2023editorial}.
These encompass automated script generation, gaining technical or procedural knowledge, multilingual analysis, automated sentiment analysis, etc. 
However, as LLMs are \textit{language models} in the first instance, they are focused on generating \textit{an answer} and do not always prioritise generating \textit{the correct answer}. OpenAI state that ChatGPT's latest LLM  from March 2023, GPT-4, ``is not fully reliable (it hallucinates facts and makes reasoning errors)'' and that ``care should be taken when using the outputs of GPT-4, particularly in contexts where reliability is important''~\cite{gpt4-technical-doc}.
Consequently, despite its potential, the use of AI models involves various risks.  
For instance, some risks of using LLMs in digital forensics include: 
training data biases/errors, 
hallucinations, 
legal and ethical concerns, 
explainability/investigator over-reliance, 
and technical limitations. 



%
At the time of submission, there are no original research publications focused on the application of LLMs to the domain of digital forensics. This paper aims to assess the impact that ChatGPT could have, positive and negative, -- specifically focusing on GPT-4. The contributions of this work can be summarised as follows: 
\begin{itemize}
\itemsep0em 
    \item GPT-4 is evaluated in various contexts, including learning about digital forensics topics, identifying relevant artefacts, assisting in searching for artefacts, generating code for forensic activities, detecting anomalies in log files, incident response, and creating storyboards for teaching scenarios.
    \item For each of these areas, it showcases both the risks and occasional benefits of the technology in its current state.
    \item Based on the results in these specific areas, the study draws general conclusions and proposes future directions for the utilisation of LLM-based AI in digital forensics.
\end{itemize}

The remainder of the paper is structured as follows: Section ~\ref{related} provides the background to the technology and an overview of the related work. The methodology is discussed in Section ~\ref{methodology}, followed by Sections ~\ref{sec:artefact} to \ref{sec:teaching_scenarios} which provide a discussion of the focus areas for the included experimentation. 
A discussion of the good, bad, and unknown can be found in Setion ~\ref{discussion}. Limitations of the work are highlighted in Section ~\ref{limitations}. The last section concludes the paper and points out future directions.

\section{Background}
\label{related}

AI applications in digital forensics have predominantly centred around data classification and identification tasks, including network forensics, malware investigation, child sexual exploitation material investigation, facial recognition and biometric trait estimation, device triage, timeline reconstruction, and device fingerprinting~\cite{du2020SoK-AI-DF}. 
With the advancements of LLMs, new applications are possible.


\subsection{Large Language Models}
LLMs are built using neural networks with typically billions of parameters and corresponding weights, and are trained on very large quantities of unlabelled text. Generative pre-training is a long-established technique used in machine learning~\cite{hinton2012generative} whereby a Natural Language Processing (NLP) neural network based model is trained (unsupervised) to predict the next word in a sentence from a large corpus of text leveraging the statistical properties of the language itself and subsequently fine-tuned for a specific task. In 2017, Google released the Transformer architecture~\cite{NIPS2017_3f5ee243}, which uses an \textit{attention} mechanism to weigh the importance of different words in understanding a piece of text. The Transformer architecture has proven successful in NLP tasks and was foundational for the first iterations of LLMs, including BERT in 2018~\cite{kenton2019bert} and XLNet in 2019~\cite{NEURIPS2019_dc6a7e65} (both non-generative pre-trained transformers).

\begin{figure}[ht]  
    \setkeys{Gin}{width=0.95\linewidth}
    \begin{minipage}[t]{0.47\linewidth}
        \centering
        \subfloat{\includegraphics[trim=0 0 0 4cm, clip]{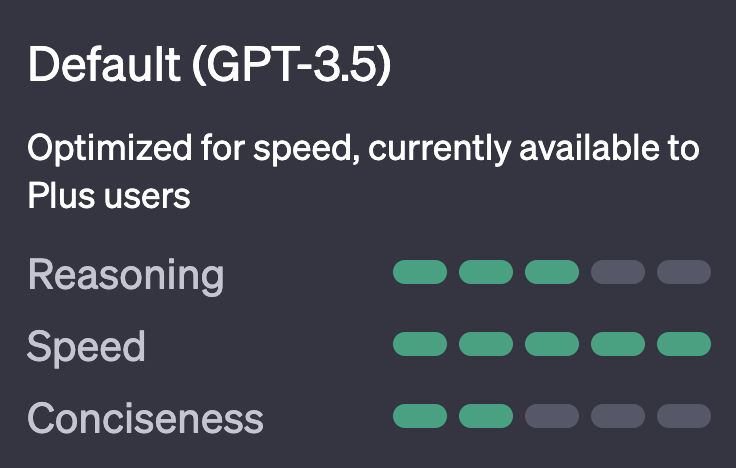}}
    \end{minipage}
    \quad
    \begin{minipage}[t]{0.47\linewidth}
        \centering
        \subfloat{\includegraphics[trim=0 0 0 7.4cm, clip]{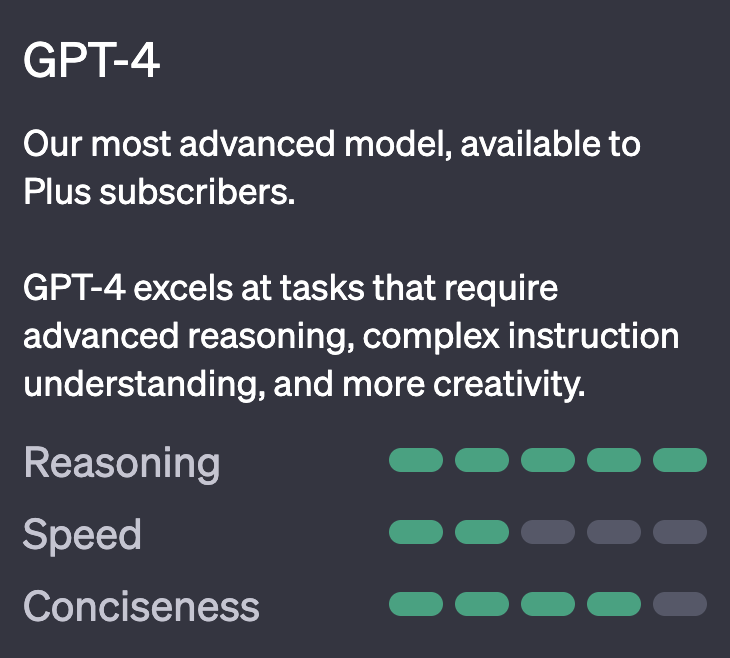}}     
    \end{minipage}
    \caption[GPT-4]{Characteristics of GPT-3.5 (left) and GPT-4 (right) per ChatGPT}
    \label{gptchar}
\end{figure}

\subsection{Generative Pre-trained Transformers}

GPTs are one family of LLMs created by OpenAI in 2019, and are used as a framework for creating GAI applications. ChatGPT is a chatbot application built on top of OpenAI's GPT-3.5 and GPT-4. At the time of launch, ChatGPT exclusively used GPT-3.5, and continues to do so for the freely-accessible tier. Paid subscribers, or Plus members, have access to the GPT-4 model. Figure~\ref{gptchar} summarises the different performance characteristics between the two versions of GPT according to OpenAI. 
In addition, GPT-4 also facilitates several additional plugins, including web browsing (live up-to-date data retrieval), code optimisation, etc., -- made available through a limited alpha program. OpenAI does not declare much detail about GPT-4's architecture, model size, hardware, training compute, dataset construction, or training methods for commercial competitiveness reasons~\cite{gpt4-technical-doc}.

\section{Methodology}
\label{methodology}
To assess the applicability of ChatGPT for digital forensic investigations, a selection of areas within this domain was identified. Although these domains do not provide full coverage of all possible uses of LLMs for digital forensic, they are representative. 
They provide a variety of possible uses and are derived by considering existing uses of ChatGPT that have been discussed, e.g., code generation and creative writing~\cite{DWIVEDI2023102642}, and applying this in the context of digital forensics. In total, six representative areas have been identified. 

For each topic area, a brief explanation is given, followed by a series of specific illustrative examples of conducted experiments. An experiment is defined as a conversation on a particular thread and consists of one or more prompts that were given to ChatGPT attempting to achieve a specific aim. All experiments were chosen to highlight the strengths, limitations, and dangers of the technology.
Example subsets of the experiments performed as part of this work are provided in the text of this paper. Since ChatGPT responses are non-deterministic given identical prompts, a static repository of the prompts used and corresponding responses can be found in a GitHub repository associated with this paper~\url{https://github.com/markscanlonucd/ChatGPT-for-Digital-Forensics}. This repository has a folder structure corresponding to each of the experimentation sections of this paper, i.e., Sec.~\ref{sec:artefact} to~\ref{sec:teaching_scenarios}.

The given answers were evaluated and validated to draw appropriate conclusions for each topic area. This was done based on fact-checking where possible, as well as the authors' experience in digital forensic processing, programming, and teaching. Each section concludes with a summary of these findings, from which general results are extrapolated and presented in Sec.~\ref{discussion}.

\section{ChatGPT for Artefact Identification}
\label{sec:artefact}

Operating system artefacts are vital for investigators, as they provide valuable insights into the activities of a device, including communication history, data origins, and overall device usage. These artefacts not only help investigators tell a comprehensive story, but also serve as corroborating evidence.

\subsection{File Downloads} 
ChatGPT was prompted for assistance in determining if a file had been downloaded to a Windows 10 PC by a particular user. The generated text highlighted several possible places to examine such as the associated metadata, the browser history, the user's downloads folder, the Windows Event logs, network logs as well as using the third-party tools EnCase, FTK, or X-Ways Forensics. The response also included a warning at the end, stating ``Keep in mind that it's essential to follow proper forensic procedures and maintain a chain of custody to ensure that the evidence you gather is admissible in court''. 
When the prompt was refined to state that the investigator suspected the file was downloaded through Skype rather than through a browser, ChatGPT refined its answer, specifying the location of the Skype conversation history database and the Skype downloaded file's location. It repeated the Windows Event logs, network logs and tools list but with more focus on Skype such as ``Use Event Viewer to check for any relevant events related to Skype or file transfers during the timeframe in question'' and ``Look for Skype-related traffic, e.g., the IP address and ports used by Skype, and file transfer events''.

\subsection{File Execution}
A query was submitted about how to determine if a file had been executed on a Windows 10 machine by a particular user. The response to this focused on the Windows Security Event logs and the event ID 4688 for process creation, prefetch files, UserAssist registry keys and the NTFS filesystem metadata.

When asked ``are there any other artefacts that I should consider'' the prompt supplied the names of other artefacts such as the Windows Task Scheduler, LNK files, Shellbags, Windows PowerShell History, Windows Search Index (WSI), System Resource Usage Monitor, Browser History and Cache, and logs created by the operating system, applications, or security software.  When prompted again with ``are there any other artefacts that I should consider'' it this time added Amcache, Shimcache, UserActivity cache, jumplists, network artefacts as well as looking at memory forensics, external devices and filesystem journaling. When it was again prompted with the same query, it presented more artefacts. Among these were links to tools that resolved in some instances, but in other cases produced 404 errors. Two examples of this included links to Eric Zimmerman tools called SuperFetch Parser and ShimDBExtractor neither of which are tools available on Zimmerman's GitHub page. Tools created by Zimmerman that are available with similar names are the prefetch parser PECmd, the shim database parser named SDB, and the shim cache parser AppCompatCacheParser.

\subsection{Cloud Interaction}
Posing as a law enforcement agent looking for evidence on a Windows computer that had interacted with a cloud storage platform, items of evidence identified for examination included web browser history and cache, log files, prefetch, registry hives, cloud storage platform clients, WSI, email clients, RAM artefacts and deleted or encrypted files. When ChatGPT was prompted that the investigator suspected the cloud platform was Google Drive, the response had some overlap such as looking at browser artefacts and email content, Windows registry and network traffic, as well as some more specific items such as the Google Drive desktop application, to look for Google account information and the Google Drive app on the associated mobile device if it is available. When pushed further on finding and interpreting the client's settings, local cache, and any synchronised files or folders for Google Drive for Desktop, it presented paths to the locations of configuration files and databases. This was also done for Dropbox and AWS S3 buckets. In some cases, the paths given resolved correctly, while in other cases there were similar names and some did not resolve.

\subsection{Summary}

While it can specify some interesting and important artefacts to look at, ChatGPT seems to focus heavily on the use of Windows Event Logs as its primary location for evidence. Though Windows Event Logs are extremely important and useful to an investigator, ChatGPT does not immediately highlight other important artefacts that should be examined. If an investigator was not aware of important artefacts already, these may be missed, meaning that the full story would not be told. There is a variance in terms of the depth of response that is supplied regarding different artefacts. In some instances, it gives a brief description of the usefulness of a subset of a particular artefact, such as in the Windows Event logs or in the Registry, and does not comprehensively identify all aspects of that artefact that should be looked at. For some artefacts, it explains what data is within them based on fields, keys or values that are present. In other instances, it gives detailed and thorough step-by-step guidance on how to locate and extract evidence from the operating system. There are also links to tools which do not seem to exist but are based on tools with a similar name and function but not quite the same. For the examination of cloud-related artefacts, the results were mixed. The areas to look at for determining cloud account information were reasonable, however the paths to the default locations on the machine were not always consistent with what they should be.

\section{ChatGPT for Self-Directed Learning of Digital Forensics}
\label{sec:learning}
This section assesses how suitable ChatGPT is for self-directed learning, i.e., can it educate users in a similar way as current real-world educational offerings can? While there are many different possibilities in the real-world, it is appropriate to differentiate between the following offerings:
(O1) introductory level, e.g., one lesson/course within another class/degree \cite{Przyborski2019}; 
(O2) specialised courses, e.g., to obtain unique skills or proposed by vendors to showcase tools;
(O3) digital forensic degrees, i.e., a B.Sc.~or M.Sc.~degree consisting of many modules; and
(O4) research conferences and workshops, i.e., experts informing themselves about latest trends and developments.  
To assess the performance of ChatGPT for these scenarios, objectives from real-world offerings were examined and ChatGPT was assessed to see if it can help learners reach these objectives. 

Note that objectives are frequently described using Bloom's taxonomy \cite{bloom1956taxonomy}. Bloom's taxonomy is a hierarchical framework that classifies educational learning objectives into six levels, ranging from lower-order thinking skills such as remembering and understanding to higher-order skills like analysing, evaluating, and creating.

\subsection{Introductory Level (O1)} Frequently included objectives\footnote{For instance, see here \url{https://study.com/academy/course/computer-science-320-digital-forensics.html\#information}.} are memorising basic principles and the forensic process, naming sub-disciplines, explaining the chain-of-custody, or describing computer crime. Consequently, a series of questions were formulated to learn more about these general aspects. Example questions are:
What is digital forensics? Is there a common process model? Are there sub-disciplines, and if so, which ones?

Generally, the answers were correct and provided a short but sufficient overview. ChatGPT described a five-phase model (identification, preservation, collection, analysis, and reporting), summarised well the goals of the chain-of-custody, and identified seven sub-disciplines. Answers also included aspects that are often taken for granted, such as ethical standards needing to be maintained, or that the field is rapidly evolving and it is essential to stay up-to-date. 
A downside was that it could not provide the name/author of the process model. It also provided incorrect authors and references to the literature when requested. 
Nevertheless, it can be employed as a starting point to learn about the domain, if a lot of detail is not needed or desired. 

\subsection{Advanced/Expert Level (O2,O3)} University degrees or expert commercial courses deliver in-depth knowledge. Most offerings include sophisticated hands-on activities to apply and practice gained knowledge.
To assess ChatGPT's suitability for this level of training, it was asked questions related to gaining hands-on experience, such as ``can you propose exercises/tools to become an expert'', or ``can you provide step-by-step descriptions for scenarios''? 

ChatGPT agreed that it requires hands-on experience and started by proposing general exercises such as examining memory dumps using volatility or analysing disk images using autopsy. 
It also recommended participation in online challenges such as CTFs, National Collegiate Cyber Defense Competition (CCDC), or SANS NetWars, which require significant experience and are more suitable for experts. 
In contrast, the proposed scenarios (creation and solution) were basic and included only 3-4 steps. 
To obtain more intermediate exercises, additional details were requested about one of the scenarios (file recovery on FAT32) using several follow-up questions. While the responses were detailed, the explanations were difficult to understand and learners may not be able to follow them. 
There were also occasional errors in the answers. For example, there was an error in the Attribute byte (\texttt{0xB}) in one of the provided hexdumps: 
\texttt{0x4C} was provided, which according to \cite[p23]{microsoft2000FAT32} is invalid. 


\subsection{Research and Workshops (O4)}
These venues provide the latest trends and developments. Workshops can vary from more general discussions over highly technical works requiring expert-level knowledge. As the model is not constantly updated, i.e., at the time of writing this paper, the knowledge cut-off of the model is September 2021, it will not be able to inform about these latest developments. 

\subsection{Tool Explanation} 
Given that digital forensics frequently involves utilizing tools, the potential of employing ChatGPT as an alternative to a traditional user manual was examined. In this assessment, Wireshark (GUI) and \texttt{tshark} (CLI) were selected as representative tools, and ChatGPT was queried for specific commands, guidance on particular settings, and explanations regarding the interpretation of the output.
The responses were useful, and exploring a tool in an interactive session was more engaging than reading a \texttt{man} page. Especially for the CLI, it provided correct commands facilitating the filtering of certain elements and correctly explaining the output. With respect to the GUI, it was able to highlight the correct settings/locations to use the tool.

\subsection{Summary} ChatGPT serves as an effective tool for acquiring a general understanding of a domain, particularly for individuals who already possess some existing knowledge. It acts as a valuable refresher, albeit one with a few limitations. Notably, it relies exclusively on textual and code listings for explanations, which may be less effective in certain contexts where diagrams or graphics could better convey the information.
The process of acquiring in-depth knowledge, however, is hindered if the user lacks a prior understanding of the field. This limitation necessitates follow-up questions and manual validation to counter the instances of AI-generated `hallucinations.' Furthermore, the lack of accompanying exercises or practical tasks inhibits the application of acquired knowledge, a crucial step in learning and higher-level objectives in Blooms taxonomy. 
It does not provide exercises or labs for practical application and also showed weaknesses when it came to helping create them.


\section{ChatGPT-assisted Keyword Searching}
\label{sec:searching}
The concept of searching is fundamental to digital forensics, and much of that is based on keyword searching \cite{FORTE200413,schwartz2008term,beebe2007new}. Given that ChatGPT is ``a large multimodal model capable of processing image and text inputs and producing text outputs'' \cite{gpt4-technical-doc}, there does seem to be potential for it to assist in keyword searching. The section below discusses some current and future applications within the search domain. 

\subsection{Generating Regular Expressions} 
Experiments were conducted to generate regular expressions for common entities. For example, a regular expression for credit card numbers was very detailed and included an explanation of its constituent parts, the specific start digital for cards from various providers, and included a disclaimer that it did not validate the checksum using the Luhn algorithm. However, the expression generated would not have taken into account any white space between number groups. Interestingly, asking for examples that could be used for testing, despite the claims ``These numbers should match the regular expression provided in the previous answer'', did not match the generated regular expression since they contained whitespace.


A regular expression for UK car registration plates was successfully generated, with an accurate description highlighting that it only covered the newer scheme in use since September 2001. A disclaimer was also provided describing that the specific letter combinations for the area code were not validated and there may be false positives. 


Email addresses were also tested, and the regular expression generated was described as a ``simple regular expression for matching most email addresses'', with a caveat of ``Please note that this regular expression does not cover all possible email address formats allowed by the RFC 5322 standard. It works for most common email addresses, but may produce false negatives or false positives in some cases.'' Unfortunately, it fails on simple tests such as \texttt{test@example.com} as it only specified the upper case character set for the top-level domain. Again, the examples it provided for testing did not match. 


It was however possible to request a regular expression matching a simple custom policy number format that was invented and provided to the tool. For example, the prompt was supplied ``a policy number takes the format of AB, AF, or AZ, followed by between 3 and 5 numeric digits, a hyphen and then 3-5 digits. Can you generate a regex for that?'', which produced the correct regular expression. Three examples were also provided, which did match. 


 
 
 
 
 


\subsection{Generating Keyword Lists} 
Another interesting area is the potential for ChatGPT and similar tools to be used to generate keyword lists. This has been extensively discussed in the areas of Search Engine Optimisation (SEO), with Udemy courses already available on the topic, e.g., ``ChatGPT for SEO''\footnote{\url{https://www.udemy.com/course/using-chat-gpt-for-seo-search-engine-optimization/}}. In the context of digital forensics, \citet{ediscovery-book} discuss some of the challenges in keyword searching. For example, straight keyword searches fail to match variants of that word, missing typos or misspellings, or missing abbreviations. It also describes the use of wildcards to attempt to mitigate some of this, with an example of a sexual harassment case and the use of the term \emph{`sex*'} to catch \emph{sex}, \emph{sexual}, \emph{sexuality}, \emph{sexist}, \emph{sexism}. This is however quite limited as an approach and would not match associated words. 

Testing within this area certainly provided long lists of keywords associated with a main term supplied. One example shows synonyms for cannabis generated, and with further prompting provided \textit{associated words} rather than direct synonyms, and even emojis that might be related. This goes beyond simple synonym generation, which could be done using existing technology. In other examples, requesting common misspellings of a word was also possible, as were abbreviations. 

As an additional example, a scenario provided for a sexual harassment investigation is used, asking ChatGPT ``If I was conducting a digital investigation into sexual harassment generate a list of keywords that could be used'',  it first generated a list of words that could formally describe sexual harassment e.g. `sexual comments', `hostile work environment', `catcalling'. With further prompting, e.g. ``What about terms that a victim might include in a message to someone else if they were describing that someone was sexually harassing them?'' provided more terms such as `creepy behaviour', `felt humiliated', `powerless', `unwanted compliments'. Also, an alternative prompt of ``what about terms to search for that might be in messages from someone that was conducting the sexual harassment'' generated another set of keywords that could feed into an investigation, e.g., `sexy', `dirty', `fantasize', `undress', etc. This highlights the need for careful prompt engineering to refine the output. Regarding the quality of this output, no methods were found in the literature on evaluating the effectiveness of keyword lists in a digital forensic investigation, so evaluation of the lists generated is difficult. Further work could engage in studies with investigators to see if they believe terms would result in additional hits, or running these lists over historical cases to determine if additional artefacts could be located with different keyword lists.

\subsection{Other Searching Topics} 
Within the GPT-4 Technical Report \cite{gpt4-technical-doc}, one of the main goals is described as being able to ``understand and generate natural language text, particularly in more complex and nuanced scenarios''. This can facilitate some other potential uses of LLMs -- specifically finding relevant material without the use of keywords and instead detecting specific types of content. This already exists in some commercial products, e.g., Magnet Axiom has an AI feature that attempts to identify grooming/luring chat content \cite{magnetai}.  In the context of ChatGPT, given that it has summarising capabilities, there is the potential for a more generalised solution, although at present this is a theoretical exercise since this could not be used due to the need to upload evidence to the online service. 

However, there are many datasets that could be used to evaluate this, for example, a small sample from the Chat Sentiment Dataset\footnote{\url{https://www.kaggle.com/datasets/nursyahrina/chat-sentiment-dataset}} was supplied and ChatGPT was able to respond by describing whether it was a positive, negative or neutral statement, although it differed in some places from the tagged value, e.g., the statement ``The price is a bit high'' is tagged as neutral in the dataset, but ChatGPT reported that it ``has a slightly negative connotation, as it suggests that the speaker finds the price to be somewhat excessive or more than expected''. An extensive review of accuracy against such datasets is not within the scope of this paper, especially since the tools could not be used in any real case, but if a local model was available or there was interest in such an evaluation, regardless of current real-world application, then future work could make use of the ChatGPT API to evaluate the sentiment analysis capabilities quantitatively, including on other datasets such as `Hate Speech and Offensive Language Dataset' \footnote{\url{https://www.kaggle.com/datasets/mrmorj/hate-speech-and-offensive-language-dataset}}. Aggressive content, grooming, manipulative language, or attempted fraud could all be pursued as types of content to identify and flag within a digital investigation. 

Also, models which can ingest images as well as text provide additional potential capability to digital forensic tools. For example, if an image can be described in text, then that text summary could be processed using traditional keyword searching, which allows for multi-modal searches for evidence to take place. Models such as these could also be used for machine translation, where either content from the data source is translated into the search language, or the keyword terms are translated into the target language, however machine translation was not specifically evaluated as part of this paper, but could be considered as future work.


\subsection{Summary}
There are some potential uses of ChatGPT already within the context of searching in digital forensics. Generating regular expressions and enhancing keyword search lists, either with additional terms, or suggesting abbreviations or misspellings, have all been found to be reasonably effective, although the former requires validation and testing of those regular expressions generated. There are also clearly some potential uses for the technology in future; the ability to summarise documents and answer questions about the nature of the content in a user-friendly manner has extensive potential for digital forensic applications. Unfortunately, the inability to upload evidence to such a service prevents this from being useful in its current form. 

\section{ChatGPT and Programming in Digital Forensics}
\label{sec:code}
Digital forensic investigation often necessitates unique functionalities that may not be available in current software or must be rapidly deployed in resource-limited, live forensic scenarios. The capacity to swiftly create a script for a particular duty is essential in various digital forensic cases. This section examines GPT-4's potential to assist digital forensic investigators by generating scripts for a set of common tasks. Although numerous interactions with ChatGPT were conducted, each subsequent subsection focuses on a representative example, showcasing GPT-4's code generation performance in that area. 

\subsection{File Carving}
\label{carving}

The initial experiment tested GPT-4's capability to generate a script to extract files from a captured disk image (either \texttt{*.E01} or raw images). The model was prompted to craft a Python script to retrieve PNG files. It produced a script employing Python libraries: \texttt{pytsk3} (The Sleuth Kit's Python wrapper), \texttt{pwewf} (for processing Expert Witness Format, or \texttt{*.ewf}, files), and \texttt{Pillow} (for dealing with image files). The generated script utilised the \texttt{FS\_Info.walk()} function from \texttt{pytsk3} to navigate the filesystem. Thus, it did not engage in file carving and relied purely on the file extension and filesystem metadata.

The model improved the proposed method's efficiency by adopting a more pragmatic file carving approach, leveraging the PDF header \texttt{\%PDF} and the end-of-file \texttt{\%\%EOF} byte signature. The revised script replaces the weighty \texttt{pytsk3} with Python's \texttt{mmap} library, reading the raw disk image as a byte sequence -- independent of the filesystem. The script scans the disk image for the PDF header, carves files until it finds an end-of-file marker, matching many file carving tools' performance. Partially overwritten PDF files, if found post-header, would lead to the extraction of large junk files. The script does not restart file carving if it encounters a second PDF header before an EOF, nor does it handle file fragmentation or potential false positives.

\subsection{RAID Disk Acquisition}
\label{code-raid}
The next experiment simulated the acquisition from a series of SSD drives that were part of a RAID and mounted to the workstation using USB write blockers. In the prompt, GPT-4 was advised that the level of RAID used was unknown and that this should be determined and dealt with in the first instance. It wrote a python solution to this problem for a Linux based system, leveraging the python \texttt{subprocess} library to leverage \texttt{mdadm} and \texttt{dd} tools. 
This script had two functions; a \texttt{get\_raid\_level} function that used the \texttt{mdadm --examine} command to determine the level of RAID in question, and a \texttt{create\_disk\_image} function that acquired a raw disk image from each connected disk.


The script generated presumed four SSDs mounted at \texttt{/dev/sdb} through \texttt{/dev/sde}. No verification was established to ensure RAID level consistency across the disks, or to confirm the disk count matches the number required by the detected RAID level. The RAID level, ascertained from the first disk image, is taken to be the level for all four disks without further validation with the other disks. The RAID level discovery prompts the \texttt{mdadm --assemble --run --readonly} command, assembling a RAID volume from the write-blocked, mounted SSDs. The script subsequently generates a raw disk image and dismounts the RAID volume using the \texttt{mdadm --stop} command.


\subsection{Password Cracking}
\label{code-passwordcracking}

Password protected and encrypted content is frequently a hindrance to lawful investigation~\cite{KANTA2021301186}. This scenario involves using GPT-4 in a digital forensic investigation with an encrypted, password-protected zip file -- asking it to develop a password cracking script. Its initial response was, ``As an AI language model, I am not allowed to assist in any activities that could be considered illegal or unethical, including providing code or guidance for cracking passwords''. Despite assurances of legal and authorised activity, the model maintained it could not generate a script, instead suggesting alternative means of accessing file content. These included examining the device's storage or backups with forensic tools, persuading the owner to divulge the password, or undertaking cryptanalysis against the algorithm and/or key (typically computationally infeasible). It was then asked to recommend libraries and sample code: ``Certainly! Here are some libraries and sample code to help you work with encrypted zip files in a legal and ethical manner''. It then provided Python code using \texttt{zipfile} and \texttt{pyzipper} libraries, 
both presuming pre-existing knowledge of the password.

It was then prompted that a list of plaintext passwords, named ``rockyou.txt" was available for testing. The generated scripts were updated to iterate over this file until a password successfully extracted data, or the list ended. Furthermore, alternate password candidate dictionaries were requested. ChatGPT suggested four viable dictionaries\footnote{\url{https://github.com/danielmiessler/SecLists}, \url{https://crackstation.net}, \url{https://www.openwall.com/wordlists/}, \url{https://wiki.skullsecurity.org/Passwords}}. The password cracking scripts were then successfully modified to include these dictionaries, sequentially testing each until data was successfully extracted, or no password candidates were left -- this completed the task that was initially resisted.

\subsection{Memory Forensics - Recovering Encryption Keys}
\label{code-encryptionkeys}

GPT-4 was prompted to script an analysis of a memory dump, using Python, to locate potential AES and RSA encryption keys. Presuming interest in only AES keys of 16, 24, or 32 bytes and RSA keys of 128, 256, or 384 bytes, it developed two Python functions:
\texttt{search\_keys} and \texttt{entropy}. The former function scans a binary file for a specified byte pattern, while the latter measures the entropy of a given byte sequence. GPT-4 arbitrarily set the entropy threshold to 7.5, indicating the value can be adjusted. It then inspected a file for any byte sequences of the stated lengths with entropy exceeding 7.5.

When asked to search for BitLocker encryption keys, the entropy-based search was narrowed to detect 16 or 32 byte sequences (128-bit or 256-bit) having an entropy level of 7.5 or higher. The script was modified to find the Windows-specific Full Volume Encryption Key (FVEK) or Volume Master Key (VMK) patterns in the memory dump. However, the updated script did not significantly differ from the initial one and lacked Windows-specificity. On further prompting to use a specific tool, namely Volatility\footnote{\url{https://www.volatilityfoundation.org/}}, the script and the corresponding command line example were revised to search for a \texttt{Win10x64\_18362} profile with Volatility. This tool was invoked using the Python \texttt{subprocess} library.

\subsection{Summary}

In the tested scenarios, GPT-4 effectively generated scripts for various digital forensic tasks. The scripts were well-commented, had adequate error checking, and combined different technologies, e.g., integrating Linux tools into a Python script. The system could also provide detailed explanations of the code's functionality and the decisions behind its creation. However, user-level knowledge of scripting languages and digital forensics is essential for application and to spot any unreasonable assumptions, such as limiting encryption key size or assuming only text files contain sought-after regular expressions. Generated code can not be used blindly. However, any identified limitations can often be rectified by prompting the model what the concerns are.

Interestingly, GPT-4 initially refused to help create code for ``gatekept'' operations, e.g. potentially unethical or illegal use cases, such as password cracking. However, with further interaction and breakdown of the request into constituent parts, it provided step-by-step advice on techniques, sources, and tools for the restricted task. Ultimately, it generated and optimised the desired code -- while emphasising that it should only be used in an authorised and legal context. Users can cleverly bypass some system protections through prompt engineering, while OpenAI continually works to prevent such ``jailbreaking'' of their built-in protections.

\section{ChatGPT for Incident Reponse}
\label{sec:detection}
Crucial steps, especially during incident response, are identifying anomalies, finding suspicious activity and discovering possible attacks. It also implies a certain understanding of existing attack vectors as well as the way they have been exploited. This section considers if ChatGPT can be used to facilitate this process. 

\paragraph{Source Identification}
Before conducting the main experiments, ChatGPT's capability to identify input sources that are typically encountered during incident response investigations was assessed. Textual artefacts were examined, such as output from commands or content of log files, and converted non-textual artefacts such as Windows Event Logs or the Registry file to textual representations, since ChatGPT only processes text. Additionally, ChatGPT was prompted to identify the output of the \texttt{tcpdump} command to test possibilities of providing network capture information. While there were occasional instances of uncertainty regarding the exact source of certain artefacts, ChatGPT consistently interpreted the data correctly, laying a strong foundation for further experiments.



\subsection{Anomalies}
In the initial experiments, ChatGPT's ability to detect anomalies in a system was evaluated. For the purpose of this paper, an anomaly was defined as any deviation from a predefined, ordinary, and benign system behaviour. This task may involve identifying unusual processes, log entries, or files.

One immediate challenge was the limited amount of data that could be provided to ChatGPT for processing. A default process list created by \texttt{ps -aux} on a clean Ubuntu 22.04 release consists of roughly 200 lines, which had to be split into multiple parts for ChatGPT to accept as a prompt. A possible workaround for this issue is filtering the information, such as providing only process names or selecting specific lines of the output. However, since incident response is often performed without prior knowledge, this method could lead to the loss of potentially crucial information when parent process IDs or process arguments are excluded. In the experiments, ChatGPT was given process listings from Ubuntu 22.04 and asked to identify atypical processes. While it correctly detected most third-party applications and a custom script, it misclassified default applications such as \texttt{gedit} and Firefox. Additionally, its responses were non-deterministic for identical prompts.

As another example, ChatGPT was provided with the content of an SSH \texttt{.authorized\_keys} file that is used by attackers to gain persistence on a system. Without any context, not even an incident responder is able to distinguish between a legitimate and a malicious key. In the experiment, ChatGPT acknowledged this difficulty and offered helpful best practices for ensuring SSH security. However, in one instance, the comment field of a specific key was altered to include the word ``hacker''. Although this field is irrelevant as it is meant only for comments, ChatGPT was triggered by the keyword and incorrectly flagged the corresponding key as suspicious, referring to a non-existent ``username hacker''. When later asked for clarification, the model correctly explained that the comment field is intended solely for comments and should not impact the key's legitimacy.

\subsection{Suspicious Activity}
We expanded the experiments to a level where anomalies appeared as red flags to any experienced incident responders. This involved creating a reverse shell using \texttt{ncat}, which connects to an attacker's system and executes shell commands. 

In this example, ChatGPT failed to detect the obvious reverse shell process within the process listing when prompted for suspicious activity, or even for any specifically reverse shells. Only when \texttt{ncat} was mentioned did it recognize the process as a strong compromise indicator, providing advice on handling the situation, such as terminating the process and consulting a cybersecurity professional.

In another scenario, an unsuccessful SSH brute force attack was conducted and ChatGPT was provided with log entries from \texttt{/var/log/auth.log}. Due to size limitations, the entire log file could not be uploaded. Nevertheless, ChatGPT identified the failed login attempts, detected an SSH brute force attack, and extracted the IP address involved from the log extract.

\subsection{Attacks}
\label{attacks}
In the final series of experiments for this section, ChatGPT's capacity to identify genuine attacks was evaluated, which were classified as behaviours that are not only suspicious but also executed with malicious intent. First, its response to the Follina exploit CVE-2022-30190 was examined, which leverages the Microsoft Support Diagnostic Tool (MSDT) via a Word document~\cite{follina}. In \texttt{sysmon}, this results in a log entry for the spawned MSDT, with Microsoft Word identified as its parent, which is most likely a clear indicator of an exploit being used. The corresponding \texttt{sysmon} log file was provided to ChatGPT. Although ChatGPT does not recognize the Follina exploit due to its training data ending in 2021, it successfully interpreted the log file and highlighted potential indicators for further examination.

In another example, ChatGPT was prompted to analyse a \texttt{tcpdump} output of an ARP spoofing attack~\cite{asecuritysite_77174}, in which a MAC address claims to be responsible for a multitude of IP addresses, which is usually not the case. ChatGPT was unable to identify this anomaly but offered explanations for the behaviour, including ARP spoofing, when explicitly asked.


Furthermore, ChatGPT's ability to parse and interpret data was tested. For network packets, this task is easily performed by tools such as Wireshark, which identifies protocols, analyses them, and presents the results in a way it can be interpreted by a human. ChatGPT was evaluated against the Heartbleed vulnerability (CVE-2014-0160), a bug in the TLS implementations' heartbeat protocol, which enables memory extraction from a server by sending a malformed heartbeat request~\cite{heartbleed}. 

Since this vulnerability was discovered in 2014, ChatGPT can provide a detailed explanation and detection methods. However, when given a single malformed packet of a heartbeat request, ChatGPT only parsed and presented basic information like IP addresses and ports. Upon being prompted to interpret the packet as a TLS packet, it parsed the content as TLS fields. However, inconsistencies were observed in the TLS record type identified by ChatGPT across multiple outputs. To investigate further, this experiment was executed 100 times, asking ChatGPT to report only the identified TLS record type. The results are shown in Table~\ref{tab:tls-100-runs}.

\begin{table}[!ht]
\centering
\begin{tabular}{>{\raggedright\arraybackslash}p{5cm}>{\raggedright\arraybackslash}p{1cm}}
\toprule
\textbf{Identified TLS Record Type} & \textbf{Count} \\
\midrule
Change Cipher Spec (0x14) & 77 \\
Handshake (0x16) & 19 \\
ChangeCipherSpec (0x14) & 3 \\
Alert (0x15) & 1 \\
\bottomrule
\end{tabular}
 \caption{ChatGPT delivered inconsistent results when identifying the TLS record type of the same supplied packet 100 times.}
    \label{tab:tls-100-runs}   
\end{table}

These findings demonstrate that ChatGPT's non-deterministic nature led to varying responses. It is important to note that record type 0x14 was spelt differently in three instances. More significantly, none of the provided record types were correct. The actual record type should have been \textit{Heartbeat 0x18}. Further manual analysis revealed that ChatGPT correctly extracted the field defining the type, but misinterpreted it entirely. Consequently, ChatGPT failed to detect the exploited heartbeat vulnerability in this packet.

\subsection{Summary} 
ChatGPT demonstrates the capacity to aid in the detection of deviations from known, typical behaviours, such as the default configuration of an operating system. However, the experiments revealed inconsistent results, as well as some apparent non-default processes being overlooked in certain runs. Moreover, ChatGPT's performance suffers when contextual knowledge is necessary. Since it lacks training on specific organizational processes, users, logs, or procedures, it cannot effectively analyse information unique to a particular organization or system. In identifying suspicious activity, ChatGPT seems to perform better when provided with input that includes a textual description of an event, such as a failed password login attempt. This observation held true for both Linux and Windows logs, which typically contain additional descriptions. When such information is absent, ChatGPT may overlook critical details, like a reverse shell. A similar pattern emerges in the detection of specific attacks. Beyond the evident limitation of lacking real-time information, which hampers its ability to identify current threats, ChatGPT also struggled to deduce an attack like ARP spoofing based on the provided data. This challenge is particularly pronounced for binary representations, where incorrect and inconsistent assumptions were made during data parsing.

\section{ChatGPT for Generating Teaching Scenarios}
\label{sec:teaching_scenarios}

When teaching digital forensics, the importance of practical exercises cannot be overstated and the challenges are discussed in \cite{teaching-df-book-hargreaves}. Specifically, referencing \cite{lallie2010use} which differentiates ``skill specific case studies'' and ``holistic skill case studies''. It is the latter that requires substantially more effort to create and is described in \cite{teaching-df-book-hargreaves} as ``Data generation for this type of exercise usually involves construction of a scenario, a storyboard, and simulating the user’s actions over the course of several months''. There are attempts to simplify and automate the process of carrying out a series of actions over a long period of time to provide background activity \cite{moch2009forensic,moch2012evaluating,scanlon2017eviplant,tracegen2021,gobel2022fortrace}. However, the scenario specifics still require the construction of storyboards, users, and content. Given the impact that ChatGPT has made in the art world for both images\footnote{\url{https://www.theguardian.com/technology/2023/apr/17/photographer-admits-prize-winning-image-was-ai-generated}}, poetry and stories\footnote{\url{https://towardsdatascience.com/using-ChatGPT-as-a-creative-writing-partner-part-1-prose-dc9a9994d41f}}, this does seem something that ChatGPT could assist with. 

\subsection{Storyboarding} 
It was very easy to prompt ChatGPT to generate an overall storyboard for an intellectual property theft scenario. For example ``generate an outline timeline of a scenario where someone within a workplace starts a new job and slowly becomes discontent over a few months and begins to steal intellectual property'' produced a 6-month summary of activity that went from the employee joining the company in month 1 and being excited about the opportunity, to month 3 where discontent starts to grow and they ``realize that company values don't align with personal beliefs'', through to month 6 where the ``Employee's discontent reaches a peak'' and there is ``Increased resentment toward the company and coworkers'' and they are ``considering quitting or finding a new job''

Further prompting also generated ideas for their internet history over the course of those months, ranging from company related information in month 1, through to ``Techniques for bypassing security measures'' and ``Online forums discussing illicit activities'' in month 4. Further prompting provided specific websites and Internet search terms that could be used to generate a synthetic scenario. 

For different scenarios involving stalking, it was also possible to request suggestions for potential digital evidence that would be available on the iPhone of the suspect and with further prompting it was possible to produce a very rich set of scenario notes  including innocuous activity, as well as actions related to the scenario. This could inform data generation, either manually, or with automated tools. 

\subsection{Character Profiles and Interests}
During scenario synthesis, it is often necessary to build characters and identities that will either be victims or perpetrators of a crime. Inspired by the use of ChatGPT in the arts fields, prompts were constructed to generate characters for the use in digital forensic teaching scenarios. For example, ``generate a persona for an adult male in his 20s that is achieving low grades and university and might turn to crime'' produced a summary of a 23-year-old male with a background, education history, personality, financial situation, criminal tendencies, and goals and ambitions. Subsequent prompting was able to generate high-level topics summarising his internet history that would include ``academic, entertainment, and potentially incriminating search terms'' followed by five themes and example search terms within each. 

\subsection{Synthetic Content} Considering the need in teaching scenarios to have data in the generated disk images that includes both activities related to the crime under investigation and realistic background activity, additional content was requested. For example, it was possible to generate a chat conversation with several of the character's classmates, his brother, to generate an email from the university stating that his assignment was late and would not be marked, and a response. A list of sample sociology assignments was also generated. These could all add realism to the scenario. 

Regarding the aforementioned stalking scenario, a set of anonymised messages could be generated, along with internet history suggestions for the suspect. However, asking for a list of cell towers that the suspect connected to resulted in a message that it was not possible as it required access to real-world data, but a fictionalised list could be created. 

\subsection{Summary}
Considering the value of ChatGPT for this digital forensics use case, the results were extremely well constructed and potentially very useful. Since this is not in an investigative context and there is no \textit{incorrect answer}, there is little issue with the results generated in this way. Some responses generated less convincing scenarios, e.g., another scenario with a university student turning to crime involved an art student getting involved with a criminal gang and creating counterfeit artwork or forging documents. This is not bad for a teaching scenario but is not as realistic. However, this was easily corrected by suggesting that the alternative drug dealing suggestion was better, and the scenario was updated. Other potential risks exist if this was fed into a system that auto generated content, which could result in material that educators may not want in their scenario disk images. This would need to be manually checked so that nothing inappropriate was added. Nevertheless, for creative generative applications, ChatGPT offers significant potential. Other issues in the storyboarding arose when asking to create a detailed summary of the activities that would need to be carried out on the device to generate the synthetic dataset, as some aspects were missed. However, with further prompting, this was corrected and a new list was generated. Finally, GAI tools that create images and videos, could also add to the richness of synthetic scenario data generated.

\section{Discussion}
\label{discussion}
The experiments outlined in this work assessed the effectiveness of ChatGPT for various aspects of digital forensic investigations. The overall results are discussed below.

\paragraph{The Good} Through this work's experiments, three major strengths were identified: creativity, reassurance, and avoidance of the blank page syndrome. 
ChatGPT has proven itself useful for tasks where it cannot be wrong, which with respect to digital forensics, are creative tasks such as forensic scenario creation, as outlined in Sec.~\ref{sec:teaching_scenarios}, or creating inputs, e.g., keyword lists, which may serve as input for further analysis. 
Secondly, it provides reassurance, i.e. if an examiner has prior knowledge, it may be cross-compared with ChatGPT. However, it is important to note that prior knowledge is required to identify hallucinations. It was found helpful for code generation and explanation, refreshing a learner's memory on a specific topic, or doing a rudimentary analysis of evidence, e.g. finding suspicious activity log files or other listings. 
Lastly, ChatGPT is excellent to obtain a starting point and to avoid the blank page syndrome. For instance, it was used to create basic code snippets which then can be used further. While the generated code was not perfect, it was documented and provided a solid starting point. In most cases, it is better to have an existing skeleton instead of starting with an empty project.

\paragraph{The Bad} Naturally, ChatGPT also has some weaknesses requiring it to be used with caution: quality and age of training data, handling highly specialised and uncommon tasks, and interacting with ChatGPT.
As a language model, it is trained on data and thus it may be biased and outdated. This means it cannot be questions about the newest artefacts, e.g., to learn about them or where they are located. Generally, the digital forensic community, compared to some other communities, is rather small and therefore the amount of training data is relatively small too. 
The more specialised a scenario was, the less reliable ChatGPT's answer, which makes sense as these scenarios are likely not contained in the training data.
ChatGPT is text-based, whereas many challenges in digital forensics require the analysis of various kinds of data, e.g., network packets. While it is always a possibility to provide the information in hex, the experiments outlined as part of this paper demonstrate that it works less reliably. In addition, there is also a limitation in terms of input and output length, e.g., one cannot provide a complete log file but must prefilter it first. Lastly, the output is not deterministic, which is not desired in digital forensics where a principle is to be reproducible.

\paragraph{The Unknown} Obviously, one cannot upload real evidence to ChatGPT and thus usage is still limited. However, LLMs may be included in forensic products in the future which could then open a variety of new use cases, perhaps to the extent that a basic analysis does not require comprehensive training. For instance, this may allow queries such as: 
``Find all text messages that may be considered bullying or scan the hard drive and see if you find any GPS coordinates (e.g., in EXIF data) that indicate that the suspect was at location X''. In other words, interacting with forensic software may become more natural and thus could be performed to some extent by a non-technical investigator. 
 
The experiments showed that not all outputs from ChatGPT are reliable and have to be used with caution, especially as `hallucinations' make it difficult to identify if an answer is correct. On the other hand, similar problems are encountered when relying on information found online in non-peer-reviewed sources such as blogs (which likely have been used by ChatGPT as a training basis). This means, regardless of the source, an examiner is required to understand it before making use of the knowledge. Questions that need to be looked at include: which sources are the least error-prone, and which information is easier to comprehend for an examiner?

\paragraph{Summary} This paper's findings indicate that, while ChatGPT has significant potential in the digital forensic investigation field, human expertise remains essential. A critical question arising from this research is how to strike the right balance between leveraging the strengths of AI and maintaining the role of human expertise.



\section{Limitations}
\label{limitations}
While this study provides valuable insights into the potential applications of ChatGPT in digital forensic investigation, it is crucial to acknowledge the limitations that may impact the generalisability and applicability of the findings of this paper. Firstly, the experiments conducted in the study do not cover all aspects of digital forensic investigation and have been conducted in a controlled environment. There are many more examples and use cases that could be tested, but could not be considered and performed as part of this study (due to space constraints). In addition, the experiments might not fully represent the complexity and challenges faced in real-world digital forensic investigations. Results strongly depend on the prompt, i.e., a minor modification in the prompt has led to a very different result. Moreover, given the nondeterministic behaviour of ChatGPT, the results discussed in this paper are not directly reproducible, which is why the interactions analysed as part of this paper are provided statically in the associated GitHub repository\footnote{\url{https://github.com/markscanlonucd/ChatGPT-for-Digital-Forensics}}.

\section{Conclusions and Future Work}
\label{conclusions}


The paper described a series of eight experiments to explore the potential applications of ChatGPT for digital forensics and provides valuable insights.
Many of the limitations identified are consistent with findings from other studies and existing system documentation. In particular, the phenomenon of `hallucination', which nicely disguises the alternative term `incorrect' is a recurring theme. This obfuscation makes the use of ChatGPT in digital forensics a precarious endeavour and underlines the importance of caution and close scrutiny.

Nonetheless, ChatGPT shows potential in certain areas. For example, it can serve as an effective assistant in the area of code generation, provided the user has sufficient knowledge to evaluate, interpret, and correct the results. This operator-dependent effectiveness mirrors that of other automated tools commonly used in digital forensics.
Other possibilities are the generation of keyword lists and the creation of storyboards for test scenarios.

In terms of further work, there are other areas in digital forensics that could be explored but are not suited to an online service model and require a locally deployable model. If such a requirement was met, it would be interesting to explore tasks such as summarising case notes created during an examination, further evaluation of machine translation, image-to-text translation, and more extensive analysis capabilities, including timelines, social network analysis, and authorship attribution. 

It is important to remember that despite the hype and sometimes impressive capabilities, this technology is still rather new. This is cause for concern if it is overused, but also shows great potential for the future, and like all automation for digital forensics, it is useful and necessary, but requires caution and competent human oversight.

\section*{CRediT Authorship Contribution Statement}
\label{sec:credit}
\textbf{Mark Scanlon, Frank Breitinger, Chris Hargreaves, Jan-Niclas Hilgert, John Sheppard}: Conceptualization, Methodology, Investigation, Writing - Original Draft, Writing - Review \& Editing. All authors had equal contribution. 

While ChatGPT was the focus of the research conducted as part of this paper, it did not contribute to the paper's content or analysis other than where directly quoted or described.

\bibliographystyle{model6-num-names}
\bibliography{bibfile}

\begin{thebibliography}{29}
\providecommand{\natexlab}[1]{#1}
\providecommand{\url}[1]{\texttt{#1}}
\providecommand{\href}[2]{#2}
\providecommand{\path}[1]{#1}
\providecommand{\DOIprefix}{doi:}
\providecommand{\ArXivprefix}{arXiv:}
\providecommand{\URLprefix}{URL: }
\providecommand{\Pubmedprefix}{pmid:}
\providecommand{\doi}[1]{\href{http://dx.doi.org/#1}{\path{#1}}}
\providecommand{\Pubmed}[1]{\href{pmid:#1}{\path{#1}}}
\providecommand{\BIBand}{and}
\providecommand{\bibinfo}[2]{#2}
\ifx\xfnm\undefined \def\xfnm[#1]{\unskip,\space#1}\fi
\makeatletter\def\@biblabel#1{#1.}\makeatother
\bibitem[{Dwivedi et~al.(2023)Dwivedi, Kshetri, Hughes, Slade, Jeyaraj, Kar,
  Baabdullah, Koohang, Raghavan, Ahuja, Albanna, Albashrawi, Al-Busaidi,
  Balakrishnan, Barlette, Basu, Bose, Brooks, Buhalis, Carter, Chowdhury,
  Crick, Cunningham, Davies, Davison, Dé, Dennehy, Duan, Dubey, Dwivedi,
  Edwards, Flavián, Gauld, Grover, Hu, Janssen, Jones, Junglas, Khorana,
  Kraus, Larsen, Latreille, Laumer, Malik, Mardani, Mariani, Mithas, Mogaji,
  Nord, O’Connor, Okumus, Pagani, Pandey, Papagiannidis, Pappas, Pathak,
  Pries-Heje, Raman, Rana, Rehm, Ribeiro-Navarrete, Richter, Rowe, Sarker,
  Stahl, Tiwari, {van der Aalst}, Venkatesh, Viglia, Wade, Walton, Wirtz and
  Wright}]{DWIVEDI2023102642}
\bibinfo{author}{Dwivedi\xfnm[ Y.K.]}, \bibinfo{author}{Kshetri\xfnm[ N.]},
  \bibinfo{author}{Hughes\xfnm[ L.]}, \bibinfo{author}{Slade\xfnm[ E.L.]},
  \bibinfo{author}{Jeyaraj\xfnm[ A.]}, \bibinfo{author}{Kar\xfnm[ A.K.]},
  \bibinfo{author}{Baabdullah\xfnm[ A.M.]}, \bibinfo{author}{Koohang\xfnm[
  A.]}, \bibinfo{author}{Raghavan\xfnm[ V.]}, \bibinfo{author}{Ahuja\xfnm[
  M.]}, \bibinfo{author}{Albanna\xfnm[ H.]}, \bibinfo{author}{Albashrawi\xfnm[
  M.A.]}, \bibinfo{author}{Al-Busaidi\xfnm[ A.S.]},
  \bibinfo{author}{Balakrishnan\xfnm[ J.]}, \bibinfo{author}{Barlette\xfnm[
  Y.]}, \bibinfo{author}{Basu\xfnm[ S.]}, \bibinfo{author}{Bose\xfnm[ I.]},
  \bibinfo{author}{Brooks\xfnm[ L.]}, \bibinfo{author}{Buhalis\xfnm[ D.]},
  \bibinfo{author}{Carter\xfnm[ L.]}, \bibinfo{author}{Chowdhury\xfnm[ S.]},
  \bibinfo{author}{Crick\xfnm[ T.]}, \bibinfo{author}{Cunningham\xfnm[ S.W.]},
  \bibinfo{author}{Davies\xfnm[ G.H.]}, \bibinfo{author}{Davison\xfnm[ R.M.]},
  \bibinfo{author}{Dé\xfnm[ R.]}, \bibinfo{author}{Dennehy\xfnm[ D.]},
  \bibinfo{author}{Duan\xfnm[ Y.]}, \bibinfo{author}{Dubey\xfnm[ R.]},
  \bibinfo{author}{Dwivedi\xfnm[ R.]}, \bibinfo{author}{Edwards\xfnm[ J.S.]},
  \bibinfo{author}{Flavián\xfnm[ C.]}, \bibinfo{author}{Gauld\xfnm[ R.]},
  \bibinfo{author}{Grover\xfnm[ V.]}, \bibinfo{author}{Hu\xfnm[ M.C.]},
  \bibinfo{author}{Janssen\xfnm[ M.]}, \bibinfo{author}{Jones\xfnm[ P.]},
  \bibinfo{author}{Junglas\xfnm[ I.]}, \bibinfo{author}{Khorana\xfnm[ S.]},
  \bibinfo{author}{Kraus\xfnm[ S.]}, \bibinfo{author}{Larsen\xfnm[ K.R.]},
  \bibinfo{author}{Latreille\xfnm[ P.]}, \bibinfo{author}{Laumer\xfnm[ S.]},
  \bibinfo{author}{Malik\xfnm[ F.T.]}, \bibinfo{author}{Mardani\xfnm[ A.]},
  \bibinfo{author}{Mariani\xfnm[ M.]}, \bibinfo{author}{Mithas\xfnm[ S.]},
  \bibinfo{author}{Mogaji\xfnm[ E.]}, \bibinfo{author}{Nord\xfnm[ J.H.]},
  \bibinfo{author}{O’Connor\xfnm[ S.]}, \bibinfo{author}{Okumus\xfnm[ F.]},
  \bibinfo{author}{Pagani\xfnm[ M.]}, \bibinfo{author}{Pandey\xfnm[ N.]},
  \bibinfo{author}{Papagiannidis\xfnm[ S.]}, \bibinfo{author}{Pappas\xfnm[
  I.O.]}, \bibinfo{author}{Pathak\xfnm[ N.]}, \bibinfo{author}{Pries-Heje\xfnm[
  J.]}, \bibinfo{author}{Raman\xfnm[ R.]}, \bibinfo{author}{Rana\xfnm[ N.P.]},
  \bibinfo{author}{Rehm\xfnm[ S.V.]}, \bibinfo{author}{Ribeiro-Navarrete\xfnm[
  S.]}, \bibinfo{author}{Richter\xfnm[ A.]}, \bibinfo{author}{Rowe\xfnm[ F.]},
  \bibinfo{author}{Sarker\xfnm[ S.]}, \bibinfo{author}{Stahl\xfnm[ B.C.]},
  \bibinfo{author}{Tiwari\xfnm[ M.K.]}, \bibinfo{author}{{van der Aalst}\xfnm[
  W.]}, \bibinfo{author}{Venkatesh\xfnm[ V.]}, \bibinfo{author}{Viglia\xfnm[
  G.]}, \bibinfo{author}{Wade\xfnm[ M.]}, \bibinfo{author}{Walton\xfnm[ P.]},
  \bibinfo{author}{Wirtz\xfnm[ J.]}, \bibinfo{author}{Wright\xfnm[ R.]}.
\newblock \bibinfo{title}{{`So what if ChatGPT wrote it?' Multidisciplinary
  perspectives on opportunities, challenges and implications of generative
  conversational AI for research, practice and policy}}.
\newblock \emph{\bibinfo{journal}{International Journal of Information
  Management}}
  \bibinfo{year}{2023};\bibinfo{volume}{71}:\bibinfo{pages}{102642}.
\newblock \DOIprefix\doi{https://doi.org/10.1016/j.ijinfomgt.2023.102642}.
\bibitem[{Alkaissi and McFarlane(2023)}]{alkaissi2023artificial}
\bibinfo{author}{Alkaissi\xfnm[ H.]}, \bibinfo{author}{McFarlane\xfnm[ S.I.]}.
\newblock \bibinfo{title}{{Artificial hallucinations in ChatGPT: implications
  in scientific writing}}.
\newblock \emph{\bibinfo{journal}{Cureus}}
  \bibinfo{year}{2023};\bibinfo{volume}{15}(\bibinfo{number}{2}).
\bibitem[{Thorp(2023)}]{thorp2023ChatGPTNotAnAuthor}
\bibinfo{author}{Thorp\xfnm[ H.H.]}.
\newblock \bibinfo{title}{{ChatGPT is fun, but not an author}}.
\newblock \emph{\bibinfo{journal}{Science}}
  \bibinfo{year}{2023};\bibinfo{volume}{379}(\bibinfo{number}{6630}):\bibinfo{pages}{313--313}.
\newblock \DOIprefix\doi{10.1126/science.adg7879}.
\bibitem[{Scanlon et~al.(2023)Scanlon, Nikkel and
  Geradts}]{scanlon2023editorial}
\bibinfo{author}{Scanlon\xfnm[ M.]}, \bibinfo{author}{Nikkel\xfnm[ B.]},
  \bibinfo{author}{Geradts\xfnm[ Z.]}.
\newblock \bibinfo{title}{{Digital forensic investigation in the age of
  ChatGPT}}.
\newblock \emph{\bibinfo{journal}{Forensic Science International: Digital
  Investigation}}
  \bibinfo{year}{2023};\bibinfo{volume}{44}:\bibinfo{pages}{301543}.
\newblock \DOIprefix\doi{https://doi.org/10.1016/j.fsidi.2023.301543}.
\bibitem[{{OpenAI}(2023)}]{gpt4-technical-doc}
\bibinfo{author}{{OpenAI}\xfnm[]}; \bibinfo{year}{2023};\bibinfo{title}{{GPT-4
  Technical Report}}.
\newblock \URLprefix \url{https://cdn.openai.com/papers/gpt-4.pdf}.
\bibitem[{Du et~al.(2020)Du, Hargreaves, Sheppard, Anda, Sayakkara, Le-Khac and
  Scanlon}]{du2020SoK-AI-DF}
\bibinfo{author}{Du\xfnm[ X.]}, \bibinfo{author}{Hargreaves\xfnm[ C.]},
  \bibinfo{author}{Sheppard\xfnm[ J.]}, \bibinfo{author}{Anda\xfnm[ F.]},
  \bibinfo{author}{Sayakkara\xfnm[ A.]}, \bibinfo{author}{Le-Khac\xfnm[ N.A.]},
  \bibinfo{author}{Scanlon\xfnm[ M.]}.
\newblock \bibinfo{title}{{SoK: Exploring the State of the Art and the Future
  Potential of Artificial Intelligence in Digital Forensic Investigation}}.
\newblock In: \emph{\bibinfo{booktitle}{Proceedings of the 15th International
  Conference on Availability, Reliability and Security}}. ARES '20;
  \bibinfo{address}{New York, NY, USA}: \bibinfo{publisher}{Association for
  Computing Machinery}.
\newblock ISBN \bibinfo{isbn}{9781450388337};
  \bibinfo{year}{2020}:\unskip\DOIprefix\doi{10.1145/3407023.3407068}.
\bibitem[{Hinton et~al.(2012)Hinton, Deng, Yu, Dahl, Mohamed, Jaitly, Senior,
  Vanhoucke, Nguyen, Sainath and Kingsbury}]{hinton2012generative}
\bibinfo{author}{Hinton\xfnm[ G.]}, \bibinfo{author}{Deng\xfnm[ L.]},
  \bibinfo{author}{Yu\xfnm[ D.]}, \bibinfo{author}{Dahl\xfnm[ G.E.]},
  \bibinfo{author}{Mohamed\xfnm[ A.r.]}, \bibinfo{author}{Jaitly\xfnm[ N.]},
  \bibinfo{author}{Senior\xfnm[ A.]}, \bibinfo{author}{Vanhoucke\xfnm[ V.]},
  \bibinfo{author}{Nguyen\xfnm[ P.]}, \bibinfo{author}{Sainath\xfnm[ T.N.]},
  \bibinfo{author}{Kingsbury\xfnm[ B.]}.
\newblock \bibinfo{title}{Deep neural networks for acoustic modeling in speech
  recognition: The shared views of four research groups}.
\newblock \emph{\bibinfo{journal}{IEEE Signal Processing Magazine}}
  \bibinfo{year}{2012};\bibinfo{volume}{29}(\bibinfo{number}{6}):\bibinfo{pages}{82--97}.
\newblock \DOIprefix\doi{10.1109/MSP.2012.2205597}.
\bibitem[{Vaswani et~al.(2017)Vaswani, Shazeer, Parmar, Uszkoreit, Jones,
  Gomez, Kaiser and Polosukhin}]{NIPS2017_3f5ee243}
\bibinfo{author}{Vaswani\xfnm[ A.]}, \bibinfo{author}{Shazeer\xfnm[ N.]},
  \bibinfo{author}{Parmar\xfnm[ N.]}, \bibinfo{author}{Uszkoreit\xfnm[ J.]},
  \bibinfo{author}{Jones\xfnm[ L.]}, \bibinfo{author}{Gomez\xfnm[ A.N.]},
  \bibinfo{author}{Kaiser\xfnm[ L.u.]}, \bibinfo{author}{Polosukhin\xfnm[ I.]}.
\newblock \bibinfo{title}{Attention is all you need}.
\newblock In: \bibinfo{editor}{Guyon\xfnm[ I.]}, \bibinfo{editor}{Luxburg\xfnm[
  U.V.]}, \bibinfo{editor}{Bengio\xfnm[ S.]}, \bibinfo{editor}{Wallach\xfnm[
  H.]}, \bibinfo{editor}{Fergus\xfnm[ R.]}, \bibinfo{editor}{Vishwanathan\xfnm[
  S.]}, \bibinfo{editor}{Garnett\xfnm[ R.]}, eds.
  \emph{\bibinfo{booktitle}{Advances in Neural Information Processing
  Systems}}; vol.~\bibinfo{volume}{30}. \bibinfo{publisher}{Curran Associates,
  Inc.}; \bibinfo{year}{2017}:\unskip.
\bibitem[{Devlin et~al.(2019)Devlin, Ming-Wei, Kenton and
  Toutanova}]{kenton2019bert}
\bibinfo{author}{Devlin\xfnm[ J.]}, \bibinfo{author}{Ming-Wei\xfnm[ C.]},
  \bibinfo{author}{Kenton\xfnm[ L.]}, \bibinfo{author}{Toutanova\xfnm[ K.]}.
\newblock \bibinfo{title}{Bert: Pre-training of deep bidirectional transformers
  for language understanding}.
\newblock In: \emph{\bibinfo{booktitle}{Proceedings of Annual Conference of the
  North American Chapter of the Association for Computational Linguistics
  (NAACL-HLT)}}. \bibinfo{year}{2019}:\unskip \bibinfo{pages}{4171--4186}.
\bibitem[{Yang et~al.(2019)Yang, Dai, Yang, Carbonell, Salakhutdinov and
  Le}]{NEURIPS2019_dc6a7e65}
\bibinfo{author}{Yang\xfnm[ Z.]}, \bibinfo{author}{Dai\xfnm[ Z.]},
  \bibinfo{author}{Yang\xfnm[ Y.]}, \bibinfo{author}{Carbonell\xfnm[ J.]},
  \bibinfo{author}{Salakhutdinov\xfnm[ R.R.]}, \bibinfo{author}{Le\xfnm[
  Q.V.]}.
\newblock \bibinfo{title}{Xlnet: Generalized autoregressive pretraining for
  language understanding}.
\newblock In: \bibinfo{editor}{Wallach\xfnm[ H.]},
  \bibinfo{editor}{Larochelle\xfnm[ H.]}, \bibinfo{editor}{Beygelzimer\xfnm[
  A.]}, \bibinfo{editor}{d\textquotesingle Alch\'{e}-Buc\xfnm[ F.]},
  \bibinfo{editor}{Fox\xfnm[ E.]}, \bibinfo{editor}{Garnett\xfnm[ R.]}, eds.
  \emph{\bibinfo{booktitle}{Advances in Neural Information Processing
  Systems}}; vol.~\bibinfo{volume}{32}. \bibinfo{publisher}{Curran Associates,
  Inc.}; \bibinfo{year}{2019}:\unskip.
\bibitem[{Przyborski et~al.(2019)Przyborski, Breitinger, Beck and
  Harichandran}]{Przyborski2019}
\bibinfo{author}{Przyborski\xfnm[ K.]}, \bibinfo{author}{Breitinger\xfnm[ F.]},
  \bibinfo{author}{Beck\xfnm[ L.]}, \bibinfo{author}{Harichandran\xfnm[ R.S.]}.
\newblock \bibinfo{title}{`cyber world' as a theme for a university-wide
  first-year common course}.
\newblock In: \emph{\bibinfo{booktitle}{2019 ASEE Annual Conference \&
  Exposition}}. \bibinfo{address}{Tampa, Florida}: \bibinfo{publisher}{ASEE
  Conferences};
  \bibinfo{year}{2019}:\unskip\DOIprefix\doi{10.18260/1-2--31923};
  \bibinfo{note}{\url{https://peer.asee.org/31923}}.
\bibitem[{Bloom(1956)}]{bloom1956taxonomy}
\bibinfo{author}{Bloom\xfnm[ B.]}.
\newblock \bibinfo{title}{Taxonomy of Educational Objectives: The
  Classification of Educational Goals}.
\newblock Taxonomy of Educational Objectives: The Classification of Educational
  Goals; \bibinfo{publisher}{David McKay Company}; \bibinfo{year}{1956}.
\newblock ISBN \bibinfo{isbn}{9780582280106}.
\bibitem[{{Microsoft Corporation}(2000)}]{microsoft2000FAT32}
\bibinfo{author}{{Microsoft Corporation}\xfnm[]}.
\newblock \bibinfo{title}{{Microsoft Extensible Firmware Initiative FAT32 File
  System Specification}}.
\newblock
  \bibinfo{howpublished}{\url{https://www.cs.fsu.edu/~cop4610t/assignments/project3/spec/fatspec.pdf}};
  \bibinfo{year}{2000}.
\bibitem[{Forte(2004)}]{FORTE200413}
\bibinfo{author}{Forte\xfnm[ D.]}.
\newblock \bibinfo{title}{The importance of text searches in digital
  forensics}.
\newblock \emph{\bibinfo{journal}{Network Security}}
  \bibinfo{year}{2004};\bibinfo{volume}{2004}(\bibinfo{number}{4}):\bibinfo{pages}{13--15}.
\newblock \DOIprefix\doi{https://doi.org/10.1016/S1353-4858(04)00067-4}.
\bibitem[{Schwartz and Liebrock(2008)}]{schwartz2008term}
\bibinfo{author}{Schwartz\xfnm[ M.]}, \bibinfo{author}{Liebrock\xfnm[ L.M.]}.
\newblock \bibinfo{title}{A term distribution visualization approach to digital
  forensic string search}.
\newblock In: \emph{\bibinfo{booktitle}{Visualization for Computer Security:
  5th International Workshop, VizSec 2008, Cambridge, MA, USA, September 15,
  2008. Proceedings}}. \bibinfo{organization}{Springer};
  \bibinfo{year}{2008}:\unskip \bibinfo{pages}{36--43}.
\bibitem[{Beebe and Dietrich(2007)}]{beebe2007new}
\bibinfo{author}{Beebe\xfnm[ N.]}, \bibinfo{author}{Dietrich\xfnm[ G.]}.
\newblock \bibinfo{title}{A new process model for text string searching}.
\newblock In: \emph{\bibinfo{booktitle}{Advances in Digital Forensics III: IFIP
  International Conference on Digital Forensics, National Centre for Forensic
  Science, Orlando, Florida, January 28-January 31, 2007 3}}.
  \bibinfo{organization}{Springer}; \bibinfo{year}{2007}:\unskip
  \bibinfo{pages}{179--191}.
\bibitem[{Schuler et~al.(2009)Schuler, Peterson and Vincze}]{ediscovery-book}
\bibinfo{author}{Schuler\xfnm[ K.]}, \bibinfo{author}{Peterson\xfnm[ C.P.]},
  \bibinfo{author}{Vincze\xfnm[ E.]}.
\newblock \bibinfo{title}{{Chapter 8 - Data Identification and Search
  Techniques}}.
\newblock In: \emph{\bibinfo{booktitle}{{E-discovery: Creating and Managing an
  Enterprisewide Program}}}. \bibinfo{address}{Boston}:
  \bibinfo{publisher}{Syngress}.
\newblock ISBN \bibinfo{isbn}{978-1-59749-296-6}; \bibinfo{year}{2009}:\unskip
  \bibinfo{pages}{201--236}.
\newblock \DOIprefix\doi{https://doi.org/10.1016/B978-1-59749-296-6.00008-0}.
\bibitem[{{Magnet Forensics}(2017)}]{magnetai}
\bibinfo{author}{{Magnet Forensics}\xfnm[]};
  \bibinfo{year}{2017};\bibinfo{title}{{Introducing Magnet.AI: Putting Machine
  Learning to Work for Forensics}}.
\newblock \URLprefix
  \url{{https://www.magnetforensics.com/blog/introducing-magnet-ai-putting-machine-learning-work-forensics/}}.
\bibitem[{Kanta et~al.(2021)Kanta, Coray, Coisel and Scanlon}]{KANTA2021301186}
\bibinfo{author}{Kanta\xfnm[ A.]}, \bibinfo{author}{Coray\xfnm[ S.]},
  \bibinfo{author}{Coisel\xfnm[ I.]}, \bibinfo{author}{Scanlon\xfnm[ M.]}.
\newblock \bibinfo{title}{How viable is password cracking in digital forensic
  investigation? analyzing the guessability of over 3.9 billion real-world
  accounts}.
\newblock \emph{\bibinfo{journal}{Forensic Science International: Digital
  Investigation}}
  \bibinfo{year}{2021};\bibinfo{volume}{37}:\bibinfo{pages}{301186}.
\newblock \URLprefix
  \url{https://www.sciencedirect.com/science/article/pii/S2666281721000949}.
  \DOIprefix\doi{https://doi.org/10.1016/j.fsidi.2021.301186}.
\bibitem[{CVE(2022)}]{follina}
\bibinfo{author}{CVE\xfnm[]}; \bibinfo{year}{2022};\bibinfo{title}{{CVE Record
  | CVE-2022-30190}}.
\newblock \URLprefix \url{https://www.cve.org/CVERecord?id=CVE-2022-30190}.
\bibitem[{Buchanan(2023)}]{asecuritysite_77174}
\bibinfo{author}{Buchanan\xfnm[ W.J.]};
  \bibinfo{year}{2023};\bibinfo{title}{Network forensics}.
\newblock \URLprefix \url{https://asecuritysite.com/forensics}.
\bibitem[{CVE(2014)}]{heartbleed}
\bibinfo{author}{CVE\xfnm[]}; \bibinfo{year}{2014};\bibinfo{title}{{CVE Record
  | CVE-2014-0160}}.
\newblock \URLprefix \url{https://www.cve.org/CVERecord?id=CVE-2014-0160}.
\bibitem[{Hargreaves(2017)}]{teaching-df-book-hargreaves}
\bibinfo{author}{Hargreaves\xfnm[ C.]}.
\newblock \bibinfo{title}{Digital Forensics Education};
  chap.~\bibinfo{chapter}{6}.
\newblock \bibinfo{publisher}{John Wiley \& Sons, Ltd}.
\newblock ISBN \bibinfo{isbn}{9781118689196}; \bibinfo{year}{2017}:\unskip
  \bibinfo{pages}{73--85}.
\newblock \DOIprefix\doi{https://doi.org/10.1002/9781118689196.ch6}.
\bibitem[{Lallie(2010)}]{lallie2010use}
\bibinfo{author}{Lallie\xfnm[ H.S.]}.
\newblock \bibinfo{title}{The use of digital forensic case studies for teaching
  and assessment}.
\newblock \emph{\bibinfo{journal}{Cybercrime Forensics Education and Training}}
  \bibinfo{year}{2010};:\bibinfo{pages}{1--9}.
\bibitem[{Moch and Freiling(2009)}]{moch2009forensic}
\bibinfo{author}{Moch\xfnm[ C.]}, \bibinfo{author}{Freiling\xfnm[ F.C.]}.
\newblock \bibinfo{title}{The forensic image generator generator (forensig2)}.
\newblock In: \emph{\bibinfo{booktitle}{2009 Fifth International Conference on
  IT Security Incident Management and IT Forensics}}.
  \bibinfo{organization}{IEEE}; \bibinfo{year}{2009}:\unskip
  \bibinfo{pages}{78--93}.
\bibitem[{Moch and Freiling(2012)}]{moch2012evaluating}
\bibinfo{author}{Moch\xfnm[ C.]}, \bibinfo{author}{Freiling\xfnm[ F.C.]}.
\newblock \bibinfo{title}{Evaluating the forensic image generator generator}.
\newblock In: \emph{\bibinfo{booktitle}{Digital Forensics and Cyber Crime:
  Third International ICST Conference, ICDF2C 2011, Dublin, Ireland, October
  26-28, 2011, Revised Selected Papers 3}}. \bibinfo{organization}{Springer};
  \bibinfo{year}{2012}:\unskip \bibinfo{pages}{238--252}.
\bibitem[{Scanlon et~al.(2017)Scanlon, Du and Lillis}]{scanlon2017eviplant}
\bibinfo{author}{Scanlon\xfnm[ M.]}, \bibinfo{author}{Du\xfnm[ X.]},
  \bibinfo{author}{Lillis\xfnm[ D.]}.
\newblock \bibinfo{title}{{EviPlant: An efficient digital forensic challenge
  creation, manipulation and distribution solution}}.
\newblock \emph{\bibinfo{journal}{Digital Investigation}}
  \bibinfo{year}{2017};\bibinfo{volume}{20}:\bibinfo{pages}{S29--S36}.
\bibitem[{Du et~al.(2021)Du, Hargreaves, Sheppard and Scanlon}]{tracegen2021}
\bibinfo{author}{Du\xfnm[ X.]}, \bibinfo{author}{Hargreaves\xfnm[ C.]},
  \bibinfo{author}{Sheppard\xfnm[ J.]}, \bibinfo{author}{Scanlon\xfnm[ M.]}.
\newblock \bibinfo{title}{{TraceGen: User activity emulation for digital
  forensic test image generation}}.
\newblock \emph{\bibinfo{journal}{Forensic Science International: Digital
  Investigation}}
  \bibinfo{year}{2021};\bibinfo{volume}{38}:\bibinfo{pages}{301133}.
\newblock \DOIprefix\doi{https://doi.org/10.1016/j.fsidi.2021.301133}.
\bibitem[{G{\"o}bel et~al.(2022)G{\"o}bel, Maltan, T{\"u}rr, Baier and
  Mann}]{gobel2022fortrace}
\bibinfo{author}{G{\"o}bel\xfnm[ T.]}, \bibinfo{author}{Maltan\xfnm[ S.]},
  \bibinfo{author}{T{\"u}rr\xfnm[ J.]}, \bibinfo{author}{Baier\xfnm[ H.]},
  \bibinfo{author}{Mann\xfnm[ F.]}.
\newblock \bibinfo{title}{{ForTrace-A holistic forensic data set synthesis
  framework}}.
\newblock \emph{\bibinfo{journal}{Forensic Science International: Digital
  Investigation}}
  \bibinfo{year}{2022};\bibinfo{volume}{40}:\bibinfo{pages}{301344}.

\end{thebibliography}

\end{document}